\documentclass[11pt]{article}

\usepackage[a4paper, margin=2cm]{geometry}
\usepackage{graphicx}
\usepackage[utf8]{inputenc}
\usepackage{hyperref}
\usepackage{amsmath}
\usepackage[hang,flushmargin]{footmisc}
\usepackage{caption}
\usepackage{float}
\usepackage[dvipsnames]{xcolor}
\usepackage{titling}
\usepackage{tabularx}
\usepackage{dirtytalk}
\usepackage{enumerate}
\usepackage[shortlabels]{enumitem}
\usepackage{authblk}
\usepackage{listings}
\usepackage{xcolor}
\hypersetup{
    colorlinks,
    linkcolor={black},
    citecolor={black},
    urlcolor={black}
}

\graphicspath{{./figures/}}

\newlist{requirements}{enumerate}{1}
\setlist[requirements]{%
  label=\textbf{R\arabic*}: ,
  resume
}

\newcommand{\explainedfigure}[3]{%
    \begin{figure}[H]%
        \centering%
        #2%
        \caption{#3}%
        \label{#1}%
    \end{figure}%
}

\newcommand{\seeref}[1]{%
    (see \autoref{#1})%
}

\newcommand{\abstractsection}[1]{%
  \par\addvspace{.5\baselineskip}
  \noindent\textbf{#1}\quad\ignorespaces%
}

\title{Capturing Requirements for a Data Annotation Tool for Intensive Care: Experimental User-Centered Design Study}

\author[1,2]{Marceli Wac, BSc}
\author[1]{Raul Santos Rodriguez, MSc PhD}
\author[1,2]{\\Chris McWilliams, BSc, MSc, PhD}
\author[2]{Christopher Bourdeaux,  BCHIR, MA, MB}
\affil[1]{Faculty of Engineering, University of Bristol}
\affil[2]{University Hospitals Bristol and Weston NHS Foundation Trust}

\date{September 2023}

\begin{document}

\maketitle

\begin{abstract}
\abstractsection{Background}

The continuously growing use of computational methods and artificial intelligence within healthcare contexts provides substantial opportunities for solutions to previously unsolvable problems.
In particular, techniques involving machine learning have been used together with routinely collected patient data to provide better tools for clinicians and improve patient outcomes.
Intensive care units (ICUs) are complex and data-rich environments where critically ill
patients frequently require continuous monitoring and multiple organ support.
The vast amounts of data routinely collected in the ICUs provide tremendous opportunities for machine learning, but their use comes with significant challenges.
While for certain tasks the data can be used directly, more complex and challenging problems may require additional input from humans.
This input can be provided through a process of data annotation, which involves expanding on the existing data by providing information that contextualises the data and makes it more useful for later use in the machine-learning pipeline.
Annotating data is a complex, time-consuming process that requires domain expertise and frequently, technical proficiency.
With the clinicians' time being an extremely limited resource and complexities associated with the nature of healthcare data (such as multi-modality, varying formats etc.), existing data annotation tools fail to provide an effective and time-efficient solution to this problem.

\abstractsection{Objective}

In this study, we aimed to investigate how clinicians from the ICUs approach the annotation task.
This investigation focused on establishing the characteristics of the annotation process in the context of clinical data and identifying the potential differences in the annotation workflow between different staff roles.
The overall goal of the study was to elicit requirements for a software tool that could facilitate an effective and time-efficient data annotation.

\abstractsection{Methods}

To capture the requirements, we conducted an experiment involving clinicians from the ICUs annotating printed sheets of data with periods of time during which weaning from mechanical ventilation takes place.
The participants were observed during the task and their actions were analysed in the context of Norman's Interaction Cycle to establish the requirements for the digital tool.
\newpage
\abstractsection{Results}

The annotation process followed a constant loop of annotation and evaluation, during which participants incrementally analysed and annotated the data. 
While no distinguishable differences were identified between how different staff roles annotate data, we observed preferences towards different methods for applying annotation which varied between different participants, as well as different admissions they were annotating.
We established 11 requirements for the digital data annotation tool for the healthcare setting.
5 requirements focused on facilitating features for effective annotation of individual admissions, 3 requirements surrounded a semi-automated approach to data annotation, 2 requirements dictated necessary design features to comply with the operational constraints for the tool and a final requirement pertained to the future use of the software in the context of machine learning.

\abstractsection{Conclusions}

We conducted a manual data annotation activity to establish the requirements for a digital data annotation tool.
In our analysis, we characterised the approach to the annotation exhibited by the clinicians from the ICUs and elicited 11 key requirements needed to facilitate effective data annotation in the healthcare context.

\end{abstract}

\section{Introduction}

Artificial Intelligence (AI) is a field concerned with leveraging computers to mimic human cognitive functions - such as problem-solving and decision-making \cite{russell_artificial_2016, bellman_introduction_1978}.
Depending on the task, this process can utilise a variety of methods and take many forms ranging from relatively simple rule-based expert systems (following if-then pathways), through regression (modelling the relation between different variables to predict their values) and clustering (grouping objects with similar properties together) to more complex systems such as artificial neural networks (computing systems imitating the anatomy of human brains capable of modelling non-linear processes) \cite{russell_artificial_2016, grosan_intelligent_2011, hopfield_neural_1982}.
Particularly complex problems may require solutions that cannot be achieved by traditional approaches, such as pre-programming the desired behaviour.
Instead, approaches such as Machine Learning (ML) form a subset of AI in which the algorithms are used to analyse large amounts of data to derive a method for computing a solution \cite{alpaydin_introduction_2020}.
In ML, the practice of arriving at a solution is called learning (or training) and the produced output is known as the machine learning model.
%
%
The heavy reliance on data in ML highlights the importance of ensuring appropriate quantity and quality of data and signifies its impact on the effectiveness of the created model \cite{jain_overview_2020}.
The continuously increasing popularity of ML and its rapid adoption rate in life sciences suggests that AI is at the forefront of bringing innovation to healthcare \cite{noorbakhsh-sabet_artificial_2019,davenport_potential_2019}.

Intensive care units (ICUs) are busy and complex healthcare environments where critically ill patients frequently require continuous monitoring and multiple organ support.
To provide care for those patients, clinicians working in the ICUs use a broad range of medical devices, such as ventilators, monitoring devices and intravenous pumps and lines amongst many others.
The information captured by those devices, as well as that entered by the ICU staff, is collected and collated in a clinical information system, which enables health practitioners to access large quantities of data routinely required as part of their job.
This data-rich nature and the direct influence of data on the provision of care provide a tremendous opportunity for the deployment of ML models in healthcare and in particular ICUs \cite{chapalain_is_2019}.

\explainedfigure{fig:annotation_diagram}{\includegraphics[width=\textwidth]{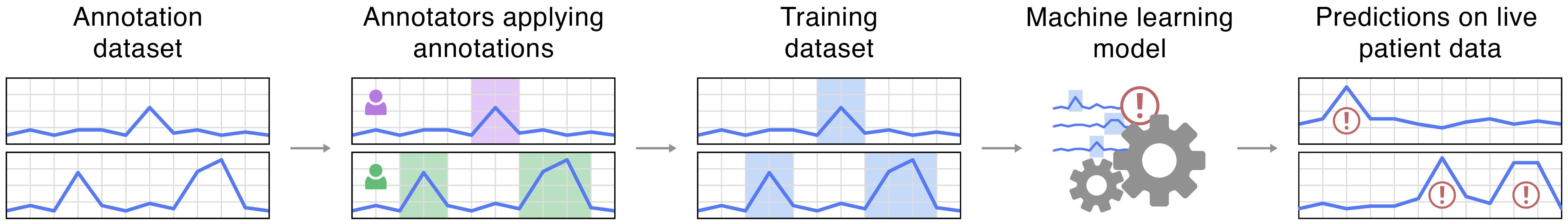}}{
Annotating data can provide additional information and context necessary to train machine learning models.}

While the data gathered in the ICUs can often be used directly, for example, to present the correlation between different vital signs and the patient prognosis, complex tasks that rely heavily on clinical experience and expertise may require human involvement to provide further guidance \cite{hunter_role_2022}.
This guidance, most frequently referred to as labels or annotations \cite{heo_artificial_2021}, can deliver additional information or context to the existing data.
An example of such a label could be a \say{yes} or \say{no} indicator of whether a patient is ready for discharge or the type and class of tumour present on an x-ray image.
With this further knowledge, ML models can take advantage of the human experience to tackle complex problems and deliver solutions that could not be inferred from the raw data alone \cite{azizi_big_2021} \seeref{fig:annotation_diagram}.
While the large volumes of data available in intensive care offer significant opportunities, working with this data also presents a unique set of challenges, as the effort required to annotate the dataset increases proportionally to its size \cite{martinez-martin_ethical_2021}.
Furthermore, the complex nature of the healthcare data and the fact that applying a single label can require clinicians to look at multiple parameters, patient history and lab results make the annotation task highly labour-intensive.
With the clinician's time being a remarkably valuable resource, ensuring the effectiveness of the data annotation workflow is paramount.
The lack of an annotation system tailored to the unique nature of the healthcare data (e.g. accessing a subset of relevant variables from a list of potentially hundreds of parameters \cite{burki_artificial_2021}) further complicates this problem.

Designing a tool for the annotation of large clinical datasets is therefore a problem that needs to be approached carefully.
The expert nature of the annotators implies that both their numbers and time are limited and should be used efficiently.
Furthermore, the clinical datasets frequently aggregate data spanning several years, resulting in a volume of data that is infeasible for direct and manual annotation.
This suggests a need for a semi-automated approach to annotation that could scale up to an entire dataset with limited input from the domain experts.

Involving end-users in the design process of the new tools prior to their development is an important aspect of designing effective software \cite{vandekerckhove_generative_2020}.
It minimises the risk of creating a system that is inefficient and helps to ensure that the developed solution meets the users' expectations and requirements \cite{kautz_participatory_2010,damodaran_user_1996}.
In the context of the data annotation software used by the experts in intensive care, this participatory design is especially critical, as each of the variety of roles (e.g. junior doctors, doctors, nurses, consultants) can generate a unique set of requirements.
Furthermore, while the staff working in the ICUs are experts in the medical domain and are trained to treat patients directly, extracting their knowledge through data annotation is an entirely different process.
For this reason, it is crucial to deploy a strategy that will ensure that the design of the tool follows a structured and systematic approach that can capture a wide variety of perspectives from its end-users.

\subsection{Objectives}

The primary objective of the research is to establish a set of criteria for the design of a data annotation tool that can be used effectively by clinicians to annotate time-series datasets from intensive care.
To achieve this goal under the limitations imposed by this setting, the system needs to account for limited access to the annotators and large volumes of data.
This suggests that a semi-automated solution should also be incorporated into the software.
Finally, to ensure the efficiency of the process, the subtleties of how clinicians approach the problem of data annotation need to be understood and accounted for in the design.
The overarching goal is, therefore, to gather requirements for a data annotation platform that is purpose-built for the intensive care data and clinicians, and as such, one that facilitates a novel way to annotate data in that domain.
The research questions defined for this project are therefore:

\subsection{Research Questions}

\begin{enumerate}[RQ1:]
    \item What are the characteristics of how clinical staff in the ICUs annotate data?
    \item How do different staff roles (nurses, junior doctors, doctors, consultants) approach the data annotation task?
    \item What are the requirements for data annotation software within the intensive care setting?
\end{enumerate}

\paragraph{RQ1: Characteristics of data annotation process}
The subtleties of how clinicians approach the data annotation task can have significant implications on the effectiveness of the annotation tool itself.
It is therefore vital to understand what characteristics of that process are specific to the time-series datasets in the intensive care setting and how they could be used to inform the design of tools that are purpose-built for experts in the clinical domain.
Obtaining these characteristics could be instrumental in developing tools that are effective at data labelling, which could result in higher-quality data and serve as a basis for deploying performant ML models in healthcare.
Understanding the novel ways in which clinical staff approach the task of data annotation within intensive care could also suggest future research directions.

\paragraph{RQ2: Inter-role differences in data annotation}
The differences between the needs of different stakeholders in the intensive care units are an essential factor to consider when designing the clinicians' tools.
By understanding the requirements of each role, the annotation tool could be designed in a way that adapts to its users' workflows, offering an effective solution that is tailored to the unique characteristics of its end-users.
In practice, this could translate to fewer frustrations associated with the use of the tool and a more productive data labelling process.

\paragraph{RQ3: Requirements for data annotation software}
The challenges associated with the clinical nature of the data and the nuances of the annotation process can dictate the approach to how such data is annotated.
Consequently, understanding these challenges and incorporating a solution that accounts for them in the design could inform the functionality of the tool and improve its effectiveness in the annotation task.
By defining the concrete requirements for the data annotation tool, the development of the platform could produce software that adequately addresses the needs of its end-users; in turn, this could result in higher-quality data annotations and machine learning models capable of tackling previously unexplored problems.

\section{Methods}


In order to understand how clinicians approach and reason about the data annotation process, first, their interaction with the annotation workflow needed to be analysed.
To achieve this, we conducted an experimental study which involved participants from ICUs manually annotating excerpts of time-series data printed on paper.
The observations made during this activity were then analysed and used as a basis for deriving a set of requirements for a digital data annotation tool.



\subsection{Task selection}

The selection of an appropriate annotation task can have implications on the approach assumed by the participants during the task, as well as the produced labels.
On the one hand, the task needs to fall within the subset that participants are familiar with, and can therefore solve with a relative degree of confidence.
Conversely, it should also pose a significant challenge in order to be representative of the annotation that would in fact be required and be complex enough to prompt an in-depth analysis of the data.
Weaning from mechanical ventilation, defined for the purpose of this study as \say{the reduction of support delivered by the ventilation with an end goal of extubation}, satisfies both of these criteria due to its challenging nature.
The process of weaning can be characterised by great variability in practice \cite{slutsky_mechanical_2012}, in which both the timing of when the weaning begins and the method in which it is delivered largely depend on the clinician in charge of the treatment \cite{slutsky_mechanical_2012, girard_efficacy_2008}.
Together with the lack of personalised guidelines \cite{nichol_what_2019} and a wide variety of ways in which patients can be weaned, weaning from mechanical ventilation constitutes a label that is difficult to capture automatically and requires the expertise of the clinicians, making it suitable for this task.

\subsection{Annotation Activity}

To facilitate the data collection, an activity was developed and conducted with participants from ICUs in University Hospitals Bristol and Weston NHS Foundation Trust (UHBW), Bristol, United Kingdom.
The activity involved manual annotation of real intensive care patient data using pens and highlighter pens on excerpts of data printed on paper sheets.
The data was formatted to resemble the clinical information system present in UHBW and displayed patient demographics in addition to a table with time-series data.
Each column of the table corresponded to an hour of the day and each row corresponded to a specific parameter; the cells of the table contained the readings of a parameter for a given hour \seeref{fig:example_annotation_sheet}.

\explainedfigure{fig:example_annotation_sheet}{\includegraphics[width=\textwidth]{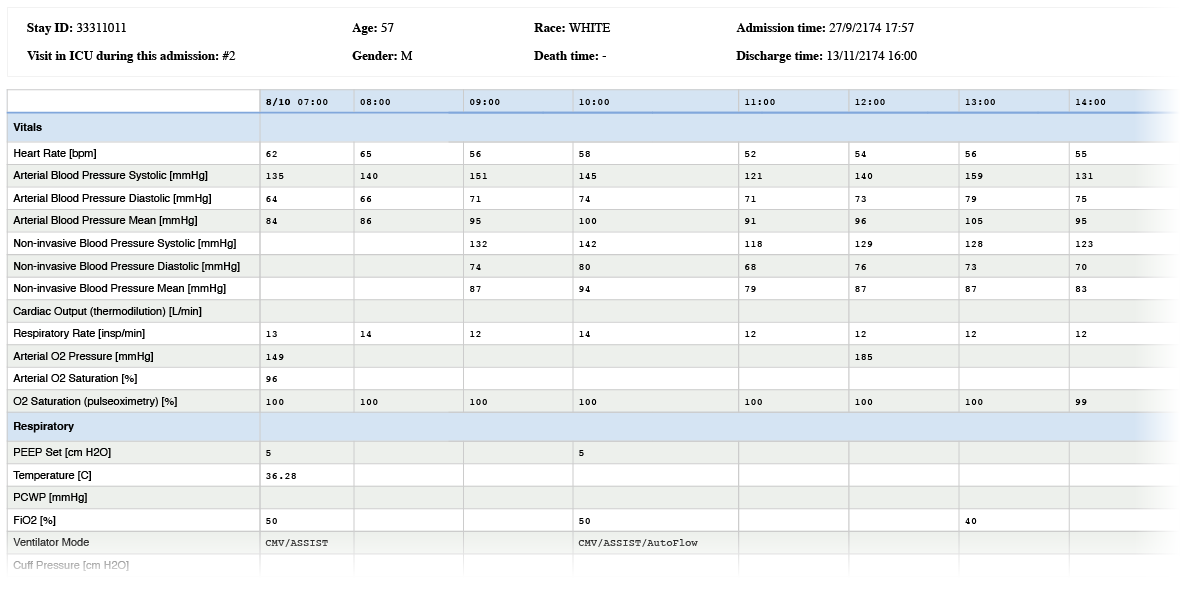}}{
Participants were asked to annotate data on the printed out sheets resembling the interface of the existing clinical system present at their ICU. The data was formatted in a table and contained parameters in each row and columns aggregating their values for each hour (depicted table was trimmed for conciseness).}

The outline brief for the activity asked participants to annotate periods of time during which weaning from mechanical ventilation takes place and allowed multiple annotations for a single admission.
Crucially, no additional information was provided as to how the data should be annotated (e.g. by circling the appropriate timestamps in the header row or commenting on the values in the table cells) and was entirely left up to participants.

\subsection{Admission data}

The data used in the activity came from a de-identified intensive care dataset called Medical Information Mart for Intensive Care IV version 2.0 (MIMIC) \cite{johnson_mimic-iv_2022}.
The criteria for eligible admissions required that subjects were at least 18 years old at the time of admission, had undergone an invasive mechanical ventilation treatment that lasted for a minimum of 24 hours, their stay in the ICU lasted for a minimum of 4 days and that their stay did not end with death, including up to 48 hours following discharge.
The time-series parameters used in the data extract were limited to those relevant to weaning from mechanical ventilation and were selected by two independent clinicians working in the ICU.
The data was extracted using a custom SQL script ran on the PostgreSQL installation of MIMIC with the concept tables computed \cite{johnson_mimic_2018}.
The script aggregated the selected parameters on an hourly basis for each eligible admission and limited it to the range surrounding the period of the mechanical ventilation treatment.

The activity was carried out in the form of a workshop held at UHBW and had each of the 7 participants from 2 distinct roles (6 junior doctors and 1 doctor) annotate 5 patient admissions.
This process was observed by an examiner who monitored and took notes on how participants approached, carried out and completed the task.
The annotated printouts were then collected for further analysis of the methods used by the participants to label the data.

\subsection{Theoritical Framework}

To analyse the observations made during this activity, the study employed a framework that could identify the challenges and opportunities in the annotation process and inform the requirements for the design of the digital tool.
Consequently, the design method for this study followed a framework defined as Norman’s Interaction Cycle (NIC), which assumes that interaction is a process of evaluation and execution between the user and the technology \cite{norman_design_2013}.
It outlines a 7-step process that begins at the goal, and through evaluation of available means and strategies, as well as the execution of these strategies - leads to achieving that goal in practice \seeref{fig:normans_interaction_cycle}.
The framework emphasizes the breakdown of the pre-production stage into planning (formulating the problem that needs to be solved), specification (outlining the strategies that can be used to tackle the problem), and performance (specifying actions that need to be taken to deploy the strategy).
Furthermore, it highlights the importance of presenting the user with feedback over three separate stages including perception (tracking the outcome of performed actions), interpretation (analysing the effect of the outcome) and comparison (evaluating whether their actions resulted in achieving the goal) \cite{norman_design_2013}.
Finally, the model aims to capture disparities between the user's intentions and actions permitted by the technology (gulf of execution) and the effort required to correctly interpret the results of their actions with regard to the desired outcome (gulf of evaluation) \cite{soegaard_gulf_2015, norman_design_2013}.

\explainedfigure{fig:normans_interaction_cycle}{\includegraphics[width=0.6\textwidth]{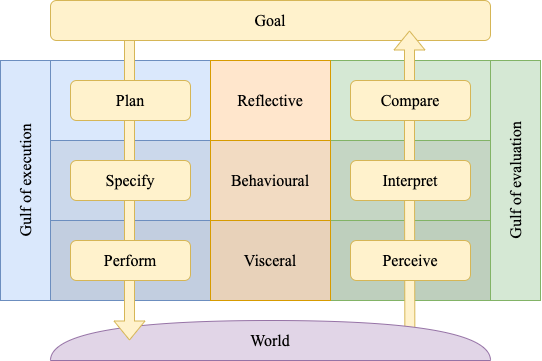}}{
Norman's Interaction Cycle follows a 7-step approach to analysing the interaction of a user with technology (adapted from \cite{norman_design_2013}).
}

\subsection{Ethics Statement}
This work was approved by the Faculty of Engineering Research Ethics Committee at the University of Bristol (case 2022-150).

\section{Results}


\subsection{Data Annotation Analysis}

In total, 7 participants took part in a workshop during which each of the participants annotated all 5 admissions that were assigned to them \seeref{fig:participants_annotating_data}.
To complete the task, participants followed a series of discrete steps when annotating data.
These steps were then separated into distinct phases of the annotation process and analysed in the context of the NIC model.

\explainedfigure{fig:participants_annotating_data}{\includegraphics[width=\textwidth]{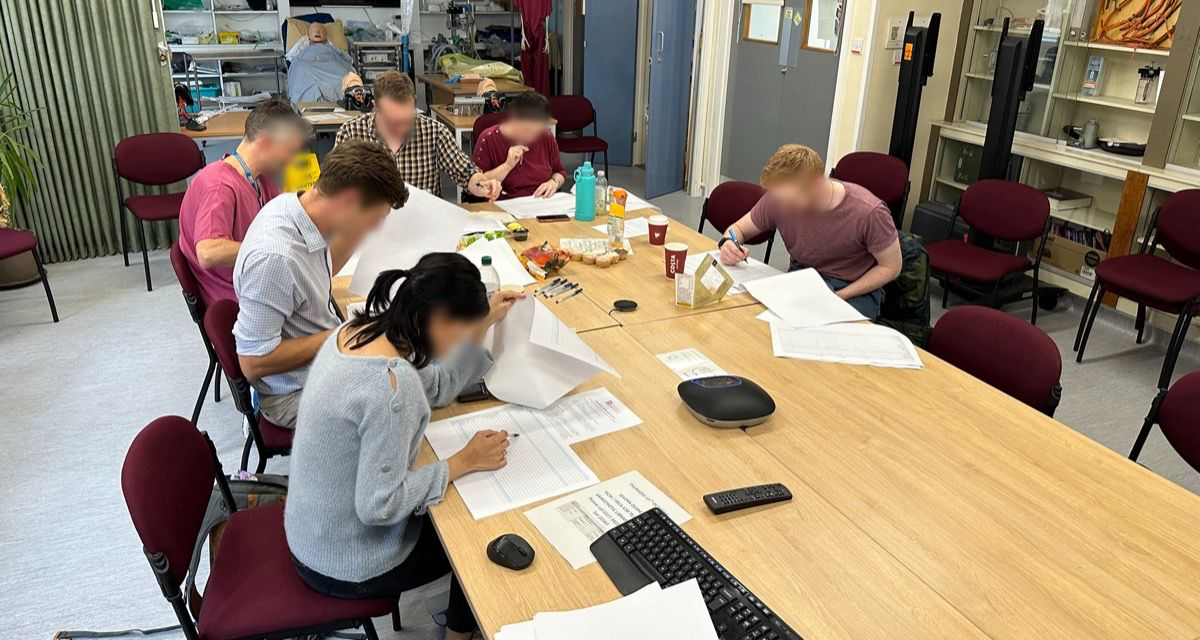}}{
Participants annotated data in a variety of ways using pen and paper during the workshop held in the ICU.
}

\paragraph{Planning}

Following the distribution of the printout sheets containing the admission data, participants began the process of familiarising themselves with the data.
This stage involved selecting a single file from the available admissions, analysing the patient demographics and browsing through the time-series data.
During the process, participants frequently cycled throughout the entire length of the time-series data from the initial admission until the discharge and back.
This suggests that during the planning stage, the primary goal of the participants was to get a high-level overview of the patient's characteristics and their stay in the ICU.
Cycling through different parts of the admission allowed to establish a broad timeline of events and the overall treatment trajectory without focusing on the minutia of its delivery.


\paragraph{Specifying}

Upon familiarisation with the specific patient admission, participants began searching for the individual points in time suggesting that weaning takes place.
While some participants preferred to start with the end of treatment where the weaning led to extubation, others analysed the data to find the point in time when weaning began.
This involved focusing on specific parameters (as suggested by following the specific rows with the pen tip) while browsing through the time-series data across different columns.
The most common parameters of interest included ventilator mode, a fraction of inspired oxygen (FiO2), set positive end-expiratory pressure (PEEP), oxygen delivery devices and flow of oxygen.
Narrowing the list of parameters to the subset significant for the task suggests the shift in focus from a broad familiarisation to a more task-oriented analysis.
It also alludes to a difference in the relevance of certain parameters to the weaning process.
Furthermore, the fact that in certain cases participants found it easier to first search for the end of weaning, rather than its start, suggests that their confidence in the accuracy of the created label (or its precise start or end) might be different across annotations.


\paragraph{Performing}

Upon finding the thresholds for either the start or end of weaning, participants performed the annotation directly on the printout sheets.
To create the labels, a variety of techniques were used which differed both between the annotators and the distinct admissions.
These included circling the start and end dates in the header row of corresponding columns, drawing vertical lines on the edges of the columns to mark the start and end periods and drawing a box around the entire dataset spanning the period of the label.
Similarly to the analysis process, some participants preferred to first annotate the end of the weaning process, rather than its start.
Together with the different ways in which the annotations were created, this suggests that participants approached the task differently and used their own preferences in analysing the admission data and creating annotations.
While the methods of annotating were different among the participants, they all shared a common set of actions of marking the beginning and end of the weaning process, even when the order of these actions was not the same.
No significant differences between how different roles annotated the data were identified among the participants of this study.

\paragraph{Perceiving}

Following the act of annotating data, participants frequently verified their labels.
This was expressed by browsing through the time-series data again and ensuring that the start and end dates corresponded to the parameter readings that suggested the annotation in the first place.
At this stage, in some cases, participants also identified the newly created labels as inaccurate or erroneous.
Further investigation of the created annotations suggests that participants felt the need to reflect on their performance immediately following the label creation.
This could also imply that the perception of the data changed when participants viewed it with the annotations overlain on top of it.



\paragraph{Interpreting}

In the cases where the created labels were erroneous, or additional insights were derived upon further investigation, participants provided supporting information to the created annotations.
These included written comments that justified the created label (e.g. \say{mode of ventilation changed from A to B} or \say{delivered oxygen reduced - indicative of weaning}) and underlined values within cells of parameters that prompted annotation.
Labels that required changing were reevaluated and amended by making necessary adjustments which usually involved crossing out the initial annotation and applying a new one.
Adding notes and comments for the created labels had important implications for the explainability of the annotations.
By circling the parameters of interest, participants provided additional context for the created label which could be used as a surrogate for their thought process and further inform the importance of specific parameters on the target label.
Furthermore, adjusting the labels following their reevaluation supports the suspected shift in perception of the data with annotations applied and could be attributed to both the learning effect and better familiarisation with the data.

\paragraph{Comparing}

Finally, upon completion of the annotation process, participants frequently sought feedback from the activity facilitator on the desired annotation count and the next steps involved in the process.
Reflecting on the annotation of an individual admission suggests that the continuity of annotation workflow spanning multiple admissions should be preserved.
It also alludes to the fact that preserving focus between the annotation of distinct admissions can have an impact on the efficiency and effectiveness of the labelling workflow as a whole.
This suggests the need for feedback on the progress of the activity and establishing a robust flow between different admissions to ensure a productive annotation process.

\subsection{User requirements}

Analysis of the annotation process observed during the workshops allowed for capturing key characteristics of how participants interacted with the task.
Together with the requirements imposed by the operational constraints and the future use of labels in the machine-learning context, these characteristics informed the user requirements for the digital annotation tool.

\subsubsection{Annotation of Individual Admissions}

The process of familiarisation with admission data during the planning and specification stages suggested that the interface of the tool should allow the end-users to conduct a similar analysis digitally.
This should include both the patient demographics and the time-series data.
Furthermore, by tracking specific values across the length of the admission participants alluded that certain parameters are more relevant to the annotation task than others and should therefore be prioritised (e.g. by placing them at the top of the list of parameters under the constraints of limited screen space).
The disparities in preferences to annotate from start to end, and vice-versa, suggested the need for flexibility in how the labels are created, which could become instrumental in making the annotation process as unobtrusive as possible.
The reevaluation of data following its annotation implied that additional insights might be derived from observing the annotated data.
As such, the digital tool should facilitate a similar ability to adjust and modify existing annotations, especially after those annotations are already displayed on top of the data.
Furthermore, this need for adjustments suggested that confidence in the accuracy of the created label can vary between different annotations.
In the context of the digital tool, deriving this information directly from the users' actions could be a complicated process.
Conversely, allowing the end-users to provide this information explicitly could result in a more accurate representation of their confidence in the created label.
The cases in which labels were further supplemented with additional information, such as the relevance of specific parameters suggested that the digital annotation system should also facilitate this.
Finally, the feedback sought upon completing the annotation indicated the need for both a progress indicator and the importance of preservation of the continuity of the annotation process on retaining participants' focus.
A digital tool could account for this by displaying the count and information about the already created labels and making it easy and efficient for participants to annotate consecutive admissions.
As such the functional requirements for the digital data annotation tool can be specified as the following:

\begin{requirements}
    \item\textbf{Data analysis} in which participants are free to navigate through the entire span of the admission data, as well as underlying patient demographics.
    \item\textbf{Label creation} functionality which is flexible enough to allow end-users to select the start and end of annotation independently and in any order.
    \item\textbf{Label adjustment} capability that facilitates an easy and convenient way to amend the created label, particularly with that label presented on top of data.
    \item\textbf{Label supplementation} feature that allows participants to provide additional context for the created annotation, including confidence in the label accuracy and relevance of different parameters.
    \item\textbf{Workflow continuity} that ensures a cohesive process of annotation and keeps the end-users engaged and focused on the task.
\end{requirements}

\subsubsection{Semi-automated annotation}

Facilitating a semi-automated approach to annotation that allows for creating labels for an entire dataset is a separate but closely related task.
As such, it can benefit from the analysis of the manual annotation process and utilise it to further inform the requirements for the digital tool.
The annotation of a single admission could be described as a bottom-up approach, in which end-users are presented with data specific to a single admission and asked to annotate it directly (optimally until the entire dataset is annotated).
Conversely, adopting an approach that automates the process would likely involve the annotation of an entire dataset based on a single user input, constituting a top-down approach.
A semi-automated approach would therefore focus on creating labels for the whole dataset without focusing on individual admissions during their creation, but potentially utilising them to evaluate the annotations and refining them upon further analysis in the context of specific admissions.
To that extent, the digital tool needs to allow the end-users to establish annotation strategies that are independent of individual admissions during their formulation.
The platform should also allow for analysis of the effectiveness of the annotation, both in the context of the entire dataset and individual admissions.

The process in which participants interacted with the individual admissions suggested several key requirements that needed to be adapted for the semi-automated approach.
The continuous loop of time-series analysis and annotation suggests that the semi-automated approach should facilitate a similar workflow in which participants can adjust and refine the annotations as they become more familiar with the data. 
Furthermore, the importance of key parameters relevant to the task should also be taken into account when designing the system to ensure its efficiency in facilitating the annotation process.
While in the context of annotating individual admissions, this information was provided explicitly by the participants, it is important to acknowledge that its presence in the data allowed them to derive further insights and adjust the labels accordingly.
The semi-automated approach should therefore also facilitate a similar functionality that provides explanations for why the label was created, allowing the end-users to further reflect on the data.
Consequently, the requirements for the digital tool facilitating a semi-automated annotation are defined as follows:

\begin{requirements}
    \item\textbf{Annotation and evaluation loop} captured as one continuous process that preserves the focus of the end-users and facilitates effective annotation.
    \item\textbf{Label analysis} of individual admissions with the overlay of created annotations on the time-series data and justification for their presence.
    \item\textbf{Dataset-wide performance metrics} that capture the effectiveness of the annotation in the form of aggregate statistics computed across the entire dataset.
\end{requirements}

\subsubsection{Operational constraints}

In addition to the requirements dictated by the participants' approach to the annotation task, there are further operational constraints that need to be reconciled to provide an effective annotation workflow.
To accommodate the busy schedules of the ICU staff, the tool should allow its end-users to access it in a way that suits their needs and does not impose strict time commitments.
As such, the end-users should be able to perform the annotation in an asynchronous way and access the platform remotely.
Furthermore, to preserve the focus and ensure that the end-users' time is used efficiently, the system should provide a responsive and time-efficient workflow that enables them to create and evaluate annotations in a short span of time.
This is particularly important for the semi-automated approach in which the timely annotation of a large volume of data is critical to preserve the continuity of the annotation and evaluation loop.
The operational requirements for the annotation tool are therefore defined as: 

\begin{requirements}
    \item\textbf{Asynchronous and remote annotation} that allows the end-users to perform the task at their convenience without further complicating their busy schedules. 
    \item\textbf{Responsiveness} of the system that provides feedback on the created annotations in a timely manner, particularly in the case of the semi-automated annotation.
\end{requirements}

\subsubsection{Labels for machine learning}

Finally, the multiple-annotator nature of the system suggests that additional metadata needs to be captured together with the created labels in order to ensure its usefulness in the machine-learning pipeline.
In particular, the process of aggregating the labels created by multiple users should account for the bias of each annotator.
Computing this bias could require an overlap in the data, allowing for a comparison of a single admission in the context of multiple annotators. 
At the same time, in the case of the single-admission annotation workflow, the fraction of the annotated dataset needs to be maximised to ensure that the limited time of the end-users is used efficiently.
Together, these requirements pose a challenge for the single-admission annotation workflow, since maximising the proportion of the annotated dataset requires a minimal overlap of admissions between different annotators.
The tradeoff between the two requirements should therefore be carefully considered when assigning admissions to annotators.

\begin{requirements}
    \item\textbf{Flexible data-splitting} solution that allows for adjustment of the data assigned to each participant and keeps track of the label authorship.
\end{requirements}

\section{Discussion}
\subsection{Principal Results}

This study sought to understand how experts from the intensive care background approach the task of data annotation in order to devise requirements for a digital tool that could be used to annotate large intensive care datasets.
Consequently, in our research, we conducted a manual annotation activity to observe how participants interacted with the annotation process and analysed their approach to elicit requirements for the design of the software.
We established 5 key requirements derived directly from the activity that could be implemented to facilitate annotation of individual patient admissions, 3 additional requirements adapted for the purpose of implementing a semi-automated annotation, 2 operational requirements that would need to be met to ensure the effectiveness of annotation process and 1 requirement stemming from the need for later use of the labels in the ML context.

\subsubsection{Approach to annotation}

The observations made during the manual annotation activity highlighted important characteristics of how experts from a clinical background approach the annotation task.
While there were no discernable differences in how participants from different roles approached the data annotation tasks within the sample of our study, we observed a variety of techniques participants used to complete the task.
The process of annotation began with an analysis of the admission data in several levels of depth.
Participants started by familiarising themselves with the admission to gain an overall picture and then proceeded to a more task-oriented analysis.
At this stage, participants focused on specific parameters, suggesting varying relevance of different parameters based on the target of the annotation.
Methods used to annotate the data differed significantly between both participants and individual admissions; in some cases, participants chose to first annotate the start of the label, while in others they began with the end.
The actual annotations were created in several ways, most commonly by enclosing the area that fell within the label or by circling the start and end dates in the header row of the table.
Following the annotation, participants reflected on the annotated data and frequently made adjustments or supplemented their labels with additional information about what prompted them to create the annotation in the first place.
Participants also reflected on their progress upon completing the annotation of each admission and sought feedback on their performance throughout the activity.

These nuances informed the requirements for the digital tool in several important ways.
Firstly, the nature of how annotators choose to analyse the data is largely preferential and might depend on the specific instance of the data. 
This suggests that the developed tool needs to provide flexibility that allows the end-users to freely create and edit annotations.
Secondly, participants frequently reevaluated data and reflected on it in the context of created annotations, which led to adjustments of the created labels.
This suggested that the data annotation process follows a continuous annotation-evaluation loop, and the digital tool should therefore facilitate a similar functionality.
It also indicated that the confidence in the accuracy of the created annotation and additional data that informed it could be valuable information which should be collected together with the labels.
Finally, the need for feedback alluded to the fact that the preservation of workflow continuity can play a key role in ensuring the productivity and efficiency of the annotation process.

\subsubsection{Design Implications}

In our study, we also found several implications of the observed approach to annotation on the design requirements for the digital annotation tool.
Observations and analysis of how participants annotated individual admissions provided transferable insights that could be adapted for the semi-automated annotation.
These included the need to facilitate the annotation-evaluation loop and preserve the continuity of the workflow, regardless of how specific annotations are collected.

The expert nature of the annotators from the clinical background suggested that providing an efficient and unobtrusive experience is critical for the effectiveness of the tool.
Together with their busy schedules, this dictated some fundamental aspects of how the software would have to be designed, such as by incorporating asynchronous and remote access to the platform.
Furthermore, the requirement to provide a time-effective annotation workflow, particularly in the context of the semi-automated approach that operates on the entire dataset, imposed additional constraints on the design requirements for the digital tool.
The feedback on the performance of the created annotations should be delivered in a timely fashion, which, in the context of large, clinical datasets, suggests the need for a solution that dynamically adapts to the volume of data.

Lastly, the use of labels in the ML workflows dictated additional requirements for how specific entries of the dataset should be assigned to individual annotators.
The conflicting characteristics of the need to maximise the fraction of the labelled dataset (thereby minimising the overlap of data) and include data shared between annotators (to account for the bias), suggested that a flexible solution that allows for customisation of the data split should also be incorporated in the design.

\subsection{Limitations and Future Work}

The activity focused on capturing the approach to data annotation was conducted with a limited number of participants, which resulted in a limited diversity of clinical roles and, potentially, perspectives from the ICU.
This suggests that some of the requirements established in this research could be particularly applicable to annotators in junior doctor positions and therefore not necessarily generalisable across entire staff working in ICUs.
The selection of the task for the manual annotation activity was made in an effort to provide an annotation experience that could be extrapolated beyond the specifics of the task.
Despite this, we acknowledge that the results it produced could be biased specifically towards the annotation of weaning from mechanical ventilation.
We acknowledge that further work in establishing the requirements for the digital tool could strengthen the understanding of how clinicians approach the data annotation task.
To that extent, we suggest that further research in this area focuses on capturing the requirements that expand beyond the confines of a single annotation task and within a broader and more diverse population from the ICUs.
Capturing a wider range of perspectives could inform the applicability of the elicited requirements and, in consequence, strengthen the resulting design of the digital tool.

The nature of the activity itself was focused strictly on the direct annotation of individual admissions, rather than the use of any assistive or automated technologies.
Although several requirements elicited in this context were applicable to the overall annotation process, including a semi-automated approach, additional requirements not captured in this study may also exist.
Further research should therefore focus on the evaluation of the proposed requirements and their utility and limitations in real-world applications.
These requirements should therefore be used to design and implement a digital data annotation platform which should be trialed within a clinical setting.
Conducting a study that investigates the feasibility of a semi-automated approach to the data annotation, particularly in comparison to the direct annotation of individual admissions could further inform the requirements for data annotation tools.

\subsection{Conclusions}

In this research, we conducted an annotation activity to capture the approach to data annotation tasks from experts from intensive care backgrounds.
We observed the participants during the task and analysed their techniques to establish key requirements for the design of a digital data annotation platform.
The requirements were then expanded by incorporating the need for a semi-automated annotation tool, the operational constraints and perspectives from the use of created labels in ML workflows.
We proposed future directions of research and suggested that a study evaluating the established requirements should be conducted to further improve the understanding of how clinicians approach the task of data annotation.

\subsection{Acknowledgements}

This work was partly supported by the Engineering and Physical Sciences Research Council Digital Health and Care Centre for Doctoral Training at the University of Bristol (UKRI grant EP/S023704/1 to MW). The work of RSR was funded by the UKRI Turing AI Fellowship (grant EP/V024817/1).
The project received an AWS Cloud Credit for Research grant.

\subsection{Author Contributions}
MW was involved in the study design, workshop facilitation and writing. RSR, CMW and CB were involved in the study design and critical revision.

\subsection{Conflicts of Interest}
None declared.

\appendix

\section{Abbreviations}
\textbf{NHS} - National Health Service \\
\textbf{UHBW} - University Hospitals Bristol and Weston NHS Foundation Trust \\
\textbf{ICU} - Intensive Care Unit \\
\textbf{NIC} - Norman's Interaction Cycle \\
\textbf{MIMIC} - Medical Information Mart for Intensive Care  \\
\textbf{SQL} - Structured Query Language \\


\newpage
\bibliographystyle{ieeetr}
\bibliography{references}

\begin{thebibliography}{10}

\bibitem{russell_artificial_2016}
S.~J. Russell and P.~Norvig, {\em Artificial intelligence: a modern approach}.
\newblock Prentice {Hall} series in artificial intelligence, Boston Columbus
  Indianapolis New York San Francisco Upper Saddle River Amsterdam Cape Town
  Dubai London Madrid Milan Munich Paris Montreal Toronto Delhi Mexico City Sao
  Paulo Sydney Hong Kong Seoul Singapore Taipei Tokyo: Pearson, third edition,
  global edition~ed., 2016.

\bibitem{bellman_introduction_1978}
R.~Bellman, {\em An {Introduction} to {Artificial} {Intelligence}: {Can}
  {Computers} {Think}?}
\newblock Boyd \& Fraser Publishing Company, 1978.
\newblock Google-Books-ID: 84xQAAAAMAAJ.

\bibitem{grosan_intelligent_2011}
C.~Grosan and A.~Abraham, {\em Intelligent systems: a modern approach}.
\newblock No.~Volume 17 in Intelligent systems reference library, Berlin:
  Springer-Verlag, 2011.

\bibitem{hopfield_neural_1982}
J.~J. Hopfield, ``Neural networks and physical systems with emergent collective
  computational abilities.,'' {\em Proceedings of the National Academy of
  Sciences}, vol.~79, pp.~2554--2558, Apr. 1982.
\newblock Publisher: Proceedings of the National Academy of Sciences.

\bibitem{alpaydin_introduction_2020}
E.~Alpaydin, {\em Introduction to machine learning}.
\newblock Adaptive computation and machine learning series, Cambridge,
  Massachusetts: The MIT Press, fourth edition~ed., 2020.

\bibitem{jain_overview_2020}
A.~Jain, H.~Patel, L.~Nagalapatti, N.~Gupta, S.~Mehta, S.~Guttula, S.~Mujumdar,
  S.~Afzal, R.~Sharma~Mittal, and V.~Munigala, ``Overview and {Importance} of
  {Data} {Quality} for {Machine} {Learning} {Tasks},'' in {\em Proceedings of
  the 26th {ACM} {SIGKDD} {International} {Conference} on {Knowledge}
  {Discovery} \& {Data} {Mining}}, (Virtual Event CA USA), pp.~3561--3562, ACM,
  Aug. 2020.

\bibitem{noorbakhsh-sabet_artificial_2019}
N.~Noorbakhsh-Sabet, R.~Zand, Y.~Zhang, and V.~Abedi, ``Artificial
  {Intelligence} {Transforms} the {Future} of {Health} {Care},'' {\em The
  American Journal of Medicine}, vol.~132, pp.~795--801, July 2019.

\bibitem{davenport_potential_2019}
T.~Davenport and R.~Kalakota, ``The potential for artificial intelligence in
  healthcare,'' {\em Future Healthcare Journal}, vol.~6, pp.~94--98, June 2019.

\bibitem{chapalain_is_2019}
X.~Chapalain and O.~Huet, ``Is artificial intelligence ({AI}) at the doorstep
  of {Intensive} {Care} {Units} ({ICU}) and operating room ({OR})?,'' {\em
  Anaesthesia Critical Care \& Pain Medicine}, vol.~38, pp.~337--338, Aug.
  2019.

\bibitem{hunter_role_2022}
B.~Hunter, S.~Hindocha, and R.~W. Lee, ``The {Role} of {Artificial}
  {Intelligence} in {Early} {Cancer} {Diagnosis},'' {\em Cancers}, vol.~14,
  p.~1524, Mar. 2022.

\bibitem{heo_artificial_2021}
M.-S. Heo, J.-E. Kim, J.-J. Hwang, S.-S. Han, J.-S. Kim, W.-J. Yi, and I.-W.
  Park, ``Artificial intelligence in oral and maxillofacial radiology: what is
  currently possible?,'' {\em Dentomaxillofacial Radiology}, vol.~50,
  p.~20200375, Mar. 2021.

\bibitem{azizi_big_2021}
S.~Azizi, B.~Mustafa, F.~Ryan, Z.~Beaver, J.~Freyberg, J.~Deaton, A.~Loh,
  A.~Karthikesalingam, S.~Kornblith, T.~Chen, V.~Natarajan, and M.~Norouzi,
  ``Big {Self}-{Supervised} {Models} {Advance} {Medical} {Image}
  {Classification},'' in {\em 2021 {IEEE}/{CVF} {International} {Conference} on
  {Computer} {Vision} ({ICCV})}, pp.~3458--3468, Oct. 2021.
\newblock ISSN: 2380-7504.

\bibitem{martinez-martin_ethical_2021}
N.~Martinez-Martin, Z.~Luo, A.~Kaushal, E.~Adeli, A.~Haque, S.~S. Kelly,
  S.~Wieten, M.~K. Cho, D.~Magnus, L.~Fei-Fei, K.~Schulman, and A.~Milstein,
  ``Ethical issues in using ambient intelligence in health-care settings,''
  {\em The Lancet Digital Health}, vol.~3, pp.~e115--e123, Feb. 2021.

\bibitem{burki_artificial_2021}
T.~K. Burki, ``Artificial intelligence hold promise in the {ICU},'' {\em The
  Lancet Respiratory Medicine}, vol.~9, pp.~826--828, Aug. 2021.

\bibitem{vandekerckhove_generative_2020}
P.~Vandekerckhove, M.~De~Mul, W.~M. Bramer, and A.~A. De~Bont, ``Generative
  {Participatory} {Design} {Methodology} to {Develop} {Electronic} {Health}
  {Interventions}: {Systematic} {Literature} {Review},'' {\em Journal of
  Medical Internet Research}, vol.~22, p.~e13780, Apr. 2020.

\bibitem{kautz_participatory_2010}
K.~Kautz, ``Participatory {Design} {Activities} and {Agile} {Software}
  {Development},'' in {\em Human {Benefit} through the {Diffusion} of
  {Information} {Systems} {Design} {Science} {Research}} (J.~Pries-Heje,
  J.~Venable, D.~Bunker, N.~L. Russo, and J.~I. DeGross, eds.), {IFIP}
  {Advances} in {Information} and {Communication} {Technology}, (Berlin,
  Heidelberg), pp.~303--316, Springer, 2010.

\bibitem{damodaran_user_1996}
L.~Damodaran, ``User involvement in the systems design process-a practical
  guide for users,'' {\em Behaviour \& Information Technology}, vol.~15,
  pp.~363--377, Jan. 1996.

\bibitem{slutsky_mechanical_2012}
A.~S. Slutsky and L.~D. Hudson, ``Mechanical {Ventilation},'' in {\em Goldman's
  {Cecil} {Medicine}}, pp.~638--645, Elsevier, 2012.

\bibitem{girard_efficacy_2008}
T.~D. Girard, J.~P. Kress, B.~D. Fuchs, J.~W.~W. Thomason, W.~D. Schweickert,
  B.~T. Pun, D.~B. Taichman, J.~G. Dunn, A.~S. Pohlman, P.~A. Kinniry, J.~C.
  Jackson, A.~E. Canonico, R.~W. Light, A.~K. Shintani, J.~L. Thompson, S.~M.
  Gordon, J.~B. Hall, R.~S. Dittus, G.~R. Bernard, and E.~W. Ely, ``Efficacy
  and safety of a paired sedation and ventilator weaning protocol for
  mechanically ventilated patients in intensive care ({Awakening} and
  {Breathing} {Controlled} trial): a randomised controlled trial,'' {\em Lancet
  (London, England)}, vol.~371, pp.~126--134, Jan. 2008.

\bibitem{nichol_what_2019}
A.~D. Nichol, P.~Geoghegan, S.~Duff, D.~J. Cooper, and D.~Devlin, ``What is the
  optimal approach to weaning and liberation from mechanical ventilation?,'' in
  {\em Evidence-based practice of critical care}, pp.~57--67.e1, Elsevier,
  Sept. 2019.

\bibitem{johnson_mimic-iv_2022}
A.~Johnson, L.~Bulgarelli, T.~Pollard, S.~Horng, L.~A. Celi, and R.~Mark,
  ``{MIMIC}-{IV},'' June 2022.

\bibitem{johnson_mimic_2018}
A.~E.~W. Johnson, D.~J. Stone, L.~A. Celi, and T.~J. Pollard, ``The {MIMIC}
  {Code} {Repository}: enabling reproducibility in critical care research,''
  {\em Journal of the American Medical Informatics Association}, vol.~25,
  pp.~32--39, Jan. 2018.

\bibitem{norman_design_2013}
D.~A. Norman, {\em The design of everyday things}.
\newblock Cambridge, MA London: The MIT Press, revised and expandes
  editons~ed., 2013.

\bibitem{soegaard_gulf_2015}
M.~Soegaard and R.~F. Dam, {\em Gulf of {Evaluation} and {Gulf} of
  {Execution}}.
\newblock Interaction Design Foundation, July 2015.

\end{thebibliography}

\end{document}